\documentclass[aps,twocolumn,preprintnumbers,amsmath,amssymb,array]{revtex4}

\usepackage{graphicx}
\usepackage{dcolumn}
\usepackage{bm}

\DeclareGraphicsExtensions{.jpg,.pdf,.eps}

\let\footnote=\endnote

\begin{document}
\title{Real-time path integral approach to nonequilibrium many-body
quantum systems}

\author{Lothar M{\"u}hlbacher and Eran Rabani}

\affiliation{School of Chemistry, The Sackler Faculty of Exact
Sciences, Tel Aviv University, Tel Aviv 69978, Israel}

\date{\today}

\begin{abstract}
A real-time path integral Monte Carlo approach is developed to study
the dynamics in a many-body quantum system until reaching a
nonequilibrium stationary state. The approach is based on augmenting
an exact reduced equation for the evolution of the system in the
interaction picture which is amenable to an efficient path integral
(worldline) Monte Carlo approach. Results obtained for a model of
inelastic tunneling spectroscopy reveal the applicability of the
approach to a wide range of physically important regimes, including
high (classical) and low (quantum) temperatures, and weak
(perturbative) and strong electron-phonon couplings.
\end{abstract}

\maketitle

In recent years there has been considerable interest in the study of
quantum mechanical systems at the nanometer scale that are driven out
of equilibrium. Experimental breakthroughs on transport in molecular
junctions have uncovered fascinating behavior in molecular systems far
from equilibrium~\cite{Reed1997,Joachim2000}. Much attention has been
paid to the study of the transport through strongly correlated systems
in which electron correlations are dominant and lead to interesting
physics such as the nonequilibrium Kondo
effect~\cite{Gordon1998,Cronenwett1998,Park2002,Liang02} and Coulomb
blockade~\cite{Elhassid00} in quantum dots/molecules, tunneling in a
Luttinger liquid~\cite{McEuen99,Yacoby02}, or inelastic effects
induced by electron-phonon interactions~\cite{Nitzan07} as probed by
inelastic electron tunneling spectroscopy~\cite{Ho98}.

Exact theoretical treatment of such many-body systems is rather sparse
and includes only a small class of simplified model
problems~\cite{Jauho1994,Schiller95,Voit96}. For a general solution
mean field equations based on the many-body nonequilibrium Green's
function (NEGF) approach can be formulated~\cite{Jauho_book} and
solved numerically. However, the mean-field NEGF approach is quite
limited since in many cases it is based on a perturbative scheme and
the inclusion of higher order corrections is not always clear or
systematic. Traditional determinant Monte Carlo (MC) approaches based
on standard discretized PIs~\cite{Sugar81,Hirsch88} have been used but
are numerically rather expensive and seem to be restriced to
imaginary-time calculations. Thus, the development of a general
approach suitable for the treatment of nonequilibrium many-body
quantum systems remains a grand challenge.

In this letter we present a novel approach aimed at obtaining exact
numerical results for various dynamical quantities such as the current
$I(t)$, conductance $g(t)$, dot population $P(t)$, etc. in a many-body
quantum system that is driven out of equilibrium. We focus on a well
studied model of inelastic tunneling
spectroscopy~\cite{Flensberg03,Galperin2004,Mitra2004,Hod06}, where a
quantum dot is coupled to two fermionic reservoirs representing the
left and right leads at chemical potentials of $\mu_{L}$ and
$\mu_{R}$, respectively, and to a bosonic bath representing the phonon
environment. Motivated by the success of real time PIMC techniques
developed for molecular chains~\cite{Muehlbacher01}, we adopt a
similar procedure and formulate an exact real time path integral (PI)
representation for the dynamical quantity of interest. To reduce the
computational complexity we integrate out the fermionic leads and the
bosonic environment and obtain expressions for their corresponding
influence functionals~\cite{Feynman63,Chen87}. We develop an adequate
Monte Carlo procedure where we propagate the density matrix from an
initial factorized arbitrary condition towards a steady-state.

A nonequilibrium quantum dot in a phonon environment can be described
by the Hamiltonian
\begin{eqnarray} \label{Hamiltonian}
\lefteqn{H
=
H_{\rm LR} + H^{(I)}_{\rm D,LR} + H_{\rm D} + H^{(I)}_{\rm D,Ph} + H_{\rm Ph}
}
\nonumber\\
&=&
\sum_{k \in L,R} \hbar\epsilon_k a^\dag_k a_k
+
\sum_{k \in L,R} \left(t_k a^\dag_k d + {\rm h.c.}\right)
+
\hbar\epsilon_D d^\dag d
\nonumber\\
&&{}+
d^\dag d \sum_\alpha M_\alpha (b^\dag_\alpha + b_\alpha)
+
\sum_\alpha \hbar\omega_\alpha (b^\dag_\alpha b_\alpha +
{}^{1\!}/_2) \,.
\end{eqnarray}
The time-dependent current from the left lead onto the dot can be
written as
\begin{eqnarray} \label{I_L}
\lefteqn{
I_L(t)
=
-e \frac{d}{dt}\langle N_L(t) \rangle
=
-\frac{2e}{\hbar} {\rm Im} \sum_{k \in L} t_k \langle a^\dag_k(t) d(t) \rangle
} \nonumber\\
&&=
-\frac{2e}{\hbar} {\rm Im} \sum_{k \in L}
 t_k e^{i\epsilon_k t}
 {\rm tr}\!\left\{W_0 U_I^\dag(t) a_k^\dag d_{H_0}(t) U_I(t) \right\} ,
\end{eqnarray}
where $W_0 = W^{(0)}_{\rm LR} \times W^{(0)}_{\rm D, Ph}$ is the
initial factorizing preparation, $U_I(t) = e^{i H_0 t/\hbar} e^{-i H
t/\hbar}$ is the interaction picture time evolution operator with $H_0
= H_{\rm LR} + H_{\rm D,Ph} = H_{\rm LR} + H_{\rm D} + H^{(I)}_{\rm
D,Ph} + H_{\rm Ph}$, and $d_{H_0}(t) = e^{iH_0t/\hbar} d\,
e^{-iH_0t/\hbar}$.

We present the approach for a quantum dot which is initially empty;
the expressions for the case of an initially occupied dot (or a mixed
preparation) can be obtained straightforwardly, as well as those for
the right current, the conductivity, the dot's population etc. After
expressing the time evolution operators in Eq.~(\ref{I_L}) by virtue
of Dyson series, $I_L(t)$ can be written as an infinite sum over
time-ordered integrals,
\begin{widetext}
\begin{equation} \label{big nasty equation}
\begin{split}
&I_L(t)
=
\frac{2e}{\hbar^2}
\sum_{n,n'=0}^\infty
\left(-\frac{1}{\hbar^2}\right)^{n+n'}
{\rm Re}
\int_0^t\!ds_{2n+1} \int_0^{s_{2n+1}}\!ds_{2n} \cdots \int_0^{s_2}\!ds_1
\int_0^t\!ds'_{2n'} \int_0^{s'_{2n'}}\!ds'_{2n'-1} \cdots \int_0^{s'_2}\!ds'_1
\\
&\qquad\qquad\qquad\qquad\times
\mathcal{L}(s_1, \dots, s_{2n+1}, t, s'_{2n'}, \dots, s'_1)\,
\mathcal{G}(s_1, \dots, s_{2n+1}, t, s'_{2n'}, \dots, s'_1)
\,,
\\
&\mathcal{L}(s_1, \dots, s_{2n+1}, t, s'_{2n'}, \dots, s'_1)
=
\sum_{\{k_j, k'_{j'}\}}
e^{i(\epsilon_{k'_1} s'_1 - \epsilon_{k'_2} s'_2
 + \ldots - \epsilon_{k'_{2n'}} s'_{2n'} + \epsilon_{k_0} t
 - \epsilon_{k_1} s_{2n+1} + \epsilon_{k_2} s_{2n}
 - \ldots - \epsilon_{k_{2n+1}} s_1 )}
\\
&\qquad\times
t_{k'_1} t^\ast_{k'_2} \cdots t^\ast_{k'_{2n'}}
t_{k_0} t^\ast_{k_1} t_{k_2} \cdots t^\ast_{k_{2n+1}}
{\rm tr}_{\rm LR}\!\left\{W^{(0)}_{\rm LR}
 a^\dag_{k'_1} a_{k'_2} a^\dag_{k'_3} \cdots
 a_{k'_{2n'}} a_{k_0}^\dag
 a_{k_1} a^\dag_{k_2} a_{k_3} \cdots a_{k_{2n+1}}
\right\} \,,
\\
&\mathcal{G}(s_1, \dots, s_{2n+1}, t, s'_{2n'}, \dots, s'_1)
=
\\
&\qquad
{\rm tr}_{\rm D,Ph} \left\{W^{(0)}_{\rm D,Ph}
 d_{H_{\rm D,Ph}}(s'_1)\, d^\dag_{H_{\rm D,Ph}}(s'_2) \cdots
 d^\dag_{H_{\rm D,Ph}}(s'_{2n'})\, d_{H_{\rm D,Ph}}(t)\,
 d^\dag_{H_{\rm D,Ph}}(s_{2n+1})\, d_{H_{\rm D,Ph}}(s_{2n})\,
 d^\dag_{H_{\rm D,Ph}}(s_1)
\right\} \,.
\end{split}
\end{equation}
\end{widetext}
Here, $d_{H_{\rm D,Ph}}(t) = e^{iH_{\rm D,Ph}t/\hbar} d\, e^{-iH_{\rm
D,Ph}t/\hbar} = d_{H_0}(t)$. The trace over the leads degrees of
freedom can now be performed exactly (see, e.g.,
Ref.~\cite{Orland_Negele}), yielding
\begin{equation} \label{L}
\mathcal{L}(t_1, \dots, t_{2N})
=
i^N \det(M) \,,
\end{equation}
where $M$ is a matrix with elements $M_{ij} = [\Sigma_{L}^{<}(t_{2j-1}
  - t_{2i}) + \Sigma_{R}^{<}(t_{2j-1} - t_{2i})
  (1-\delta_{t_{2j-1},t})(1-\delta_{t_{2i},t})]$ for $j \le i \le N$
or $[\Sigma_{L}^{>}(t_{2j-1} - t_{2i}) + \Sigma_{R}^{>}(t_{2j-1} -
  t_{2i}) (1-\delta_{t_{2j-1},t})(1-\delta_{t_{2i},t})]$ for $j > i$.
Here, $\Sigma_{L,R}^<(t)$ ($\Sigma_{L,R}^>(t)$) denotes the leads'
lesser (greater) self energy in the time domain. $\mathcal{G}$ in
Eq.~(\ref{big nasty equation}) represents a $2(n+n'+1)$ point
correlation function of the dot-phonon subsystem along the
Kadanoff-Baym contour $s: 0 \rightarrow t \rightarrow 0$. It is most
conveniently evaluated in the framework of Feynman PIs, which allows
to integrate out the phonon degrees of freedom
exactly~\cite{Feynman63}, yielding
\begin{equation} \label{G}
\mathcal{G}
=
\exp\!\left\{i\epsilon_D \int_0^t\!ds\left[\sigma(s) -
  \sigma'(s)\right]\right\} \mathcal{F}[\sigma,\sigma'] \,,
\end{equation}
where $\mathcal{F}$ denotes the Feynman-Vernon influence
functional~\cite{Feynman63} summarizing the influence of the phonons
on the dot, and $\sigma, \sigma' \in {0,1}$ denote the corresponding
forward and backward branches of the dot path, referring to the
propagations $s: 0 \rightarrow t$ and $s': t \rightarrow 0$,
respectively. Since $H_{\rm D,Ph}$ lacks any terms which could flip
the state of the dot from empty ($\sigma = 0$) to occupied ($\sigma =
1$) or vice versa, the contour-ordered sequence of time points $s_1,
\dots, s_{2n+1}, t, s'_{2n'}, \dots, s'_1$ uniquely defines the paths
$\sigma$ and $\sigma'$: Every $s^{(')}_j$ refers to one occurrence of
the dot-state altering interaction part $H^{(I)}_{\rm D,LR}$ in the
Dyson series. Therefore, the $\{s_j, s'_{j'}\}$ define the
\textit{kink} times of the dot path~\cite{Prokofev98}, between which
$\sigma(s)$ remains constant. In this spirit, Eq.~(\ref{big nasty
equation}) closely resembles the instanton expansion of the partition
function in the spin-boson model~\cite{Weissbook}.

While the expressions~(\ref{L}) and~(\ref{G}) for $\mathcal{L}$ and
$\mathcal{G}$ allow for a rather compact expression of $I_L(t)$, they
introduce retardation effects which are arbitrarily long ranged in
time, making analytical progress rather cumbersome. To allow for a
numerical evaluation of Eq.~(\ref{big nasty equation}), the integrals
in Eq.~(\ref{big nasty equation}) are approximated dividing the time
axis into $q$ discrete steps,
\begin{eqnarray}
\lefteqn{
\int_0^t\!ds_N \int_0^{s_N}\!ds_{N-1} \cdots
\int_0^{s_2}\!ds_1 f(s_1, \dots, s_N)
}
\nonumber\\
&&\simeq
\tau^N \sum_{j_N = 1}^q \sum_{j_{N-1} = 1}^{j_N - 1}
  \cdots \sum_{j_1 = 1}^{j_2 - 1} f(j_1\tau, \dots, j_N\tau) \,,
\end{eqnarray}
where $\tau = t/q$. While this introduces a systematic error of the
order of $O(\tau)$ and establishes an upper bound to the sum over the
kink numbers, the corresponding errors can be made arbitrarily small
by adjusting $q$ to correspondingly large numbers. In addition,
systematic improvement can be made by using a more accurate
integration scheme. Performing the sums over all kink numbers and the
corresponding (discretized) kink times is now equivalent to summing
over all possible (discretized) dot paths $\{\sigma_j = \sigma(j\tau),
\sigma'_{j'} = \sigma'(j'\tau)\}$, such that Eq.~(\ref{big nasty
equation}) can be written as a discretized PI,
\begin{equation} \label{MC current}
I_L(t)
=
\frac{2e}{\hbar^2} \sum_{\{\sigma_j, \sigma'_{j'}\}}
\left(-\frac{\tau}{\hbar}\right)^{n_{\rm kink}}
\mathcal{L}(\{\sigma_j, \sigma'_{j'}\})\, \mathcal{G}(\{\sigma_j,
\sigma'_{j'}\}) \,,
\end{equation}
where $n_{\rm kink}$ denotes the number of kinks of the path
$\{\sigma_j, \sigma'_{j'}\}$. Eq.~(\ref{MC current}) can now readily
be evaluated by means of PI (or worldline)
MC~\cite{Muehlbacher01,Prokofev98}.

We now turn to discuss the application of the proposed approach to the
model system described by the Hamiltonian (\ref{Hamiltonian}). In the
present approach the effect of the leads is fully determined by the
self energies $\Sigma_{L/R}^{</>}(t)$ (cf. Eq.~(\ref{L})), which are
defined in terms of $\Gamma(\epsilon) = \Gamma_L(\epsilon) +
\Gamma_R(\epsilon)$ in Fourier space~\cite{Jauho_book}.
$\Gamma(\epsilon)$ is taken to be energy independent (wide band limit)
with a soft cutoff at $\epsilon = \pm \epsilon_c$:
$\Gamma_{L/R}(\epsilon) = \frac{\Gamma_{L/R}}{[1 + e^{\nu(\epsilon -
\epsilon_c)}][1 + e^{-\nu(\epsilon + \epsilon_c)}]}$. In all results
presented below, $\Gamma_L = \Gamma_R$, $\nu = 5\Gamma = 5(\Gamma_L +
\Gamma_R)$, and $\epsilon_c = 10\Gamma$ or $20\Gamma$ to converge the
results. Similarly, the phonon influence functional $\mathcal{F}$ is
completely specified by the phonon spectral density~\cite{Feynman63},
$J(\omega) = \frac{\pi}{\hbar^2} \sum_\alpha M_{\alpha}^2
\delta(\omega - \omega_\alpha)$. For a single phonon coupled to a
secondary bath via a coupling constant $\gamma$, $J(\omega)$ becomes a
Lorentzian: $J(\omega) = \frac{\gamma \omega} {[(\omega/\omega_0)^2 -
1]^2 +
\left[\hbar^2\gamma\omega_0\omega/(2M_0^2)\right]^2}$~\cite{Leggett87}.
We note in passing that the proposed simulation scheme does neither
depend on the particular form of $\Sigma_{L/R}^{</>}(t)$ nor of
$J(\omega)$.

\begin{figure}
\begin{center}
\includegraphics[width=9cm]{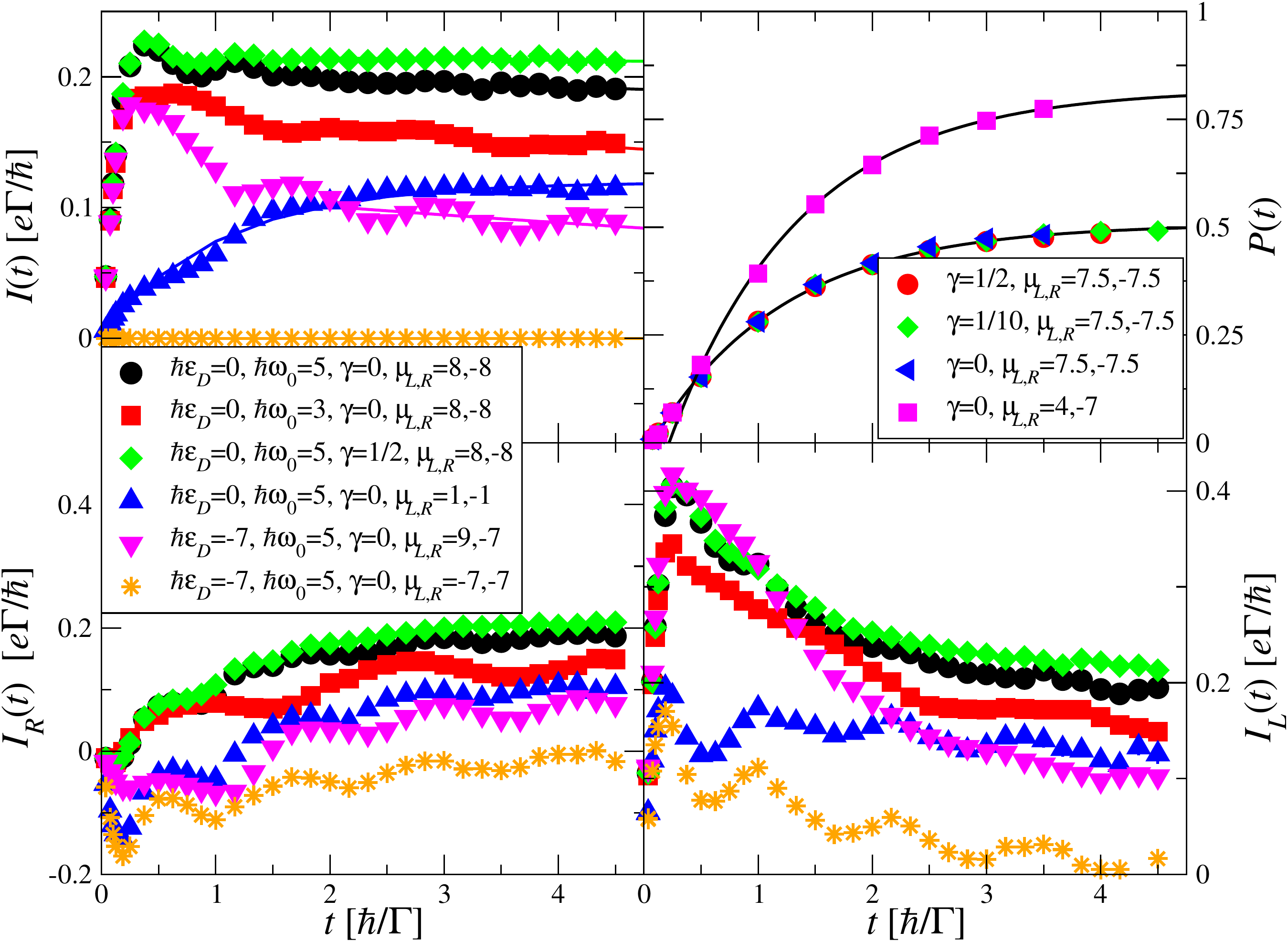}
\end{center}
\caption{Plots of the time dependent current $I_{L}(t)$ (lower left
panel), $I_{R}(t)$ (lower right panel), average current $I(t)$ (upper
left panel), and the dot population $P(t)$ (upper right panel) for
$M_{0} = 4$ and $k_B T = \frac{1}{5}$, in units
of $\Gamma=\Gamma_{L}+\Gamma_{R}$.} 
\label{fig:I_t}
\end{figure}

In Fig.~\ref{fig:I_t} we plot the time dependent current and the time
dependent dot population for different values of the model parameters
for $T<\Gamma$ (quantum regime) and $M_0>\Gamma$ (strong coupling).
Lower panels show the left ($I_L(t)$) and right ($I_R(t)$) currents
and upper panels show the average current $I(t) = \frac{1}{2}[I_L(t) +
I_R(t)]$ and the dot population $P(t)$. The left/right currents are
characterized by damped coherent oscillation with a long time
exponential decay to a steady state value. The present MC scheme
provides converged results for the time-dependent current despite the
notorious dynamical and fermionic sign problem. This can be contributed to
electronic dephasing induced by the leads and the bosonic bath, as
well as to the rather small size of the determinants in Eq.~(\ref{L})
(i.e.~half the number of kinks on the combined forward-backward dot
path) which allows for very fast MC sampling. While the steady state
current can be extracted from an exponential fit to $I_{L/R}(t)$, the
average current $I(t)$ typically decays much faster to steady state,
such that in most cases the steady state value could be obtained as
the corresponding plateau value.

\begin{figure}
\begin{center}
\includegraphics[width=8cm]{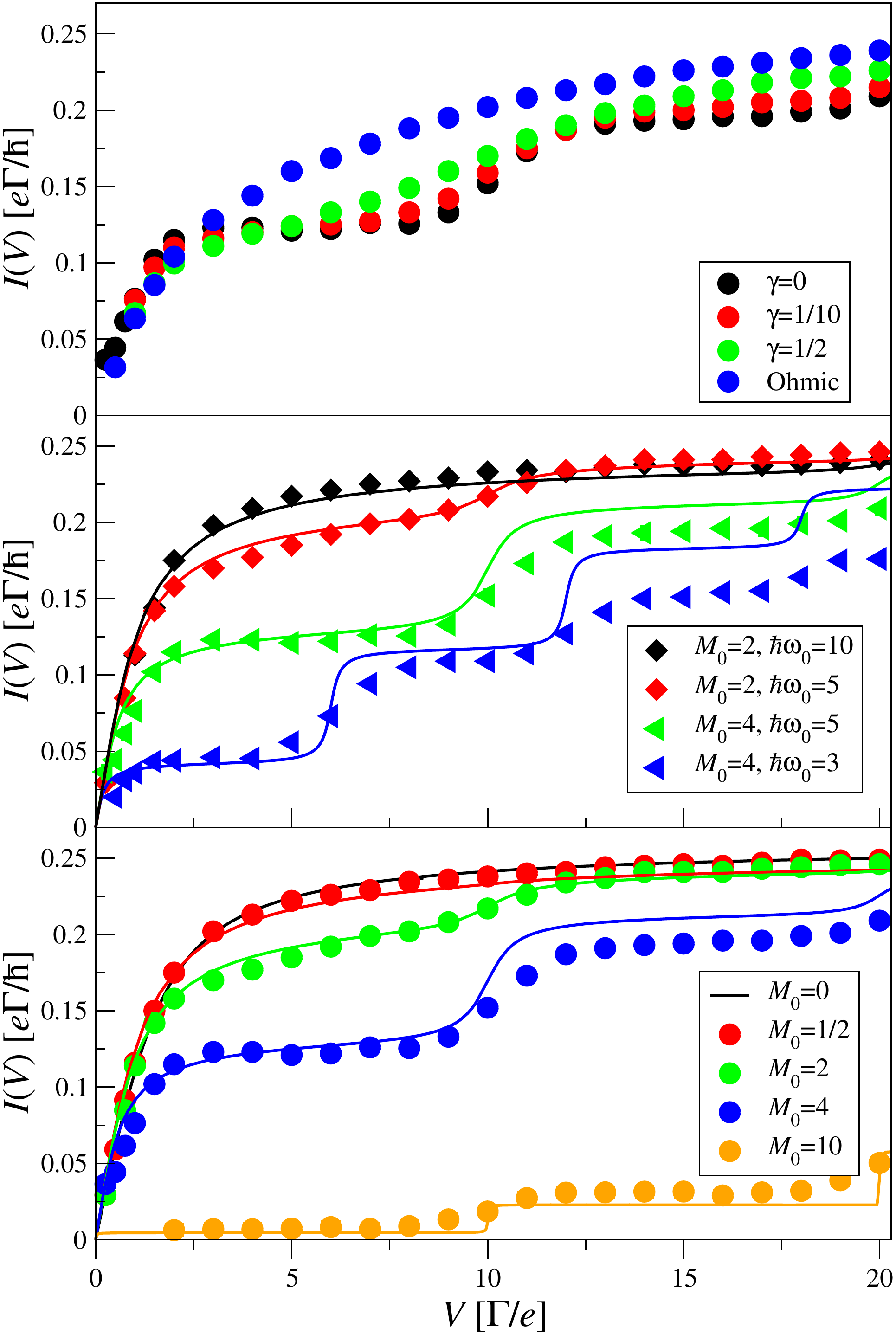}
\end{center}
\caption{Plots of the total current $I$ as a function of the bias
voltage for $\hbar \epsilon_{D}=0$, $\mu_{L,R} = \frac{eV}{2},
-\frac{eV}{2}$, and $k_BT = \frac{1}{5}$, in units of
$\Gamma=\Gamma_{L}+\Gamma_{R}$.}
\label{fig:I_V}
\end{figure}

The fact that $I(t)$ approaches faster to steady state is consistent
with other flux-based methods, \cite{Peskin03} and can be rationalized
by looking at dynamical fluctuations under equilibrium
conditions. This is depicted for the case where $\mu_L=\mu_R$ and the
system decays from an initial factorized density to equilibrium. The
left/right currents show pronounced coherent oscillations with finite
values even at $t > 5\hbar/\Gamma$ while their average value vanishes
for all times as appropriate for equilibrium conditions.
 
To analyze the limitations of the approach we have conducted a
systematic study for the case where $M_0=0$ (not shown), which
numerically is the most difficult case since decoherence effects
arising from electron-phonon coupling are neglected. Comparison of the
numerical results and the corresponding analytical solution reveals
that the present PIMC scheme provides converged results for a wide
range of gate and bias voltages, and for a wide range of temperatures
spanning the classical limit down to the quantum regime. The method is
accurate as long as $I(t)$ decays exponentially to steady state within
the time window of $t<t_{max}$. As can be seen in Fig.~\ref{fig:I_t}
the decay of $I(t)$ is slower for decreasing values of $\omega_0$ and
$\mu_L-\mu_R$. The approach is still limited for small values of these
parameters, as well as other regimes where the steady state timescale
and/or the decay time of coherent oscillations are similarily
stretched (e.g. very low temperatures); however, improvement of the MC
scheme should, in principle, overcome these limitations.

In Fig.~\ref{fig:I_V} we plot the steady state current $I$ as a
function of the bias $\mu_L-\mu_R=eV$ for different electron-phonon
couplings $M_{0}$ (lower panel), different phonon frequencies
$\omega_0$ (middle panel), and for different couplings $\gamma$ to a
secondary bath (upper panel) for $k_B T < \Gamma$ (quantum regime).
We compare the results to an approximate method based on a
generalization of the single particle approximation~\cite{Wingreen89}
to include the leading order term of the Fermi sea~\cite{Flensberg03}.

When the coupling between the primary oscillator and the secondary
phonon bath is small we observe steps in the voltage dependent current
at integer values of $eV=2\hbar\omega_0$ (middle
panel)~\cite{Nitzan07}. As the coupling to the secondary phonon bath
increases or for an Ohmic spectral density the steps diminish and
eventually disappear (upper panel), signifying the wide range of
phonon frequencies contained in the spectral density. The results for
the steady state current shown in the lower panel of
Fig.~\ref{fig:I_V} span a wide range of electron-phonon couplings from
the Landauer inelastic case, through the perturbative regime to the
strong coupling limit. As $M_{0}$ increases the value of the current
decreases from the Landauer inelastic single-channel value to lower
values. Our approach clearly captures elastic effects at all values of
$M_{0} > 0$ as depicted by the lower values of the current and by the
steps at twice the frequency of the primary phonon mode.

Comparing the exact numerical real-time PIMC results to the
approximate expansion method we find that agreement is quantitative
for small values of the voltage corresponding to the first step of
$I(V)$. For larger values we observe significant deviations between
the exact numerical results and the approximate method for
$M_{0}/\Gamma \ge 4$. These deviations signify the importance of high
order effects of the Fermi sea to the transport process, which are
naturally included in the real-time PIMC approach.

In conclusion, we have developed a novel real-time PIMC approach to
study the dynamics in open quantum systems that are driven out of
equilibrium. The approach is based on expressing the time evolution by
virtue of Dyson series, before reducing the dynamics of the entire
system by integrating out the fermionic/bosonic baths and introducing
exact influence functionals. The remaining infinite sum over
contour-ordered time integrals is then evaluated by PI (worldline) MC.
We have applied the approach to study the time-dependent current in a
well-studied model of inelastic tunneling spectroscopy, where a
quantum dot is coupled to two fermionic leads and to a bosonic phonon
bath. Numerical results indicate that the approach is robust and can
be used for a wide range of model parameters spanning the classical to
quantum limits, a range of experimentally accessible chemical bias,
different phonon frequencies, weak to strong electron-phonon
couplings, and a wide range of couplings between the primary phonon
mode and a secondary phonon bath. Furthermore, the approach provides
real-time information for various quantities, including the current,
conductance, electronic population, spectral function, and more.

We believe that our approach is capable of resolving several
shortcomings found in currently used approaches. First, it is not
based on any perturbative treatment and in principle can provide exact
numerical results. Second, it yields a compact expression for the
dynamics which can be evaluated by the proposed PI worldline method in
a numerically very efficient way. Furthermore, the real-time
propagation scheme allows to study transient phenomena and timescales
and also to include time dependent fields. In addition, since it is
based on a MC procedure, an enhancement of the accuracy can be
obtained by improving the sampling scheme, as well as by including
already existing schemes to sooth the dynamical and fermionic sign
problem. Finally, it can be applied to a general many body problem, as
long as a stable PI worldline approach can be derived.

We would like to thank Andrei Komnik, Guy Cohen, and Abe Nitzan for
stimulating discussions. This work was supported by the Israel
Ministry of Science (grant to ER). LM would like to thank the Minerva
Foundation for financial support.

\bibliographystyle{prsty} \bibliography{nonequi-pi}

\begin{thebibliography}{10}

\bibitem{Reed1997}
M.~A. Reed {\it et~al.}, Science {\bf 278},  252  (1997).

\bibitem{Joachim2000}
C. Joachim, J. k.~Gimzewski, and A. Aviram, Nature {\bf 408},  541  (2000).

\bibitem{Gordon1998}
D. Goldhaber-Gordon {\it et~al.}, Nature {\bf 391},  156  (1998).

\bibitem{Cronenwett1998}
S.~M. Cronenwett, T.~H. Oosterkamp, and L.~P. Kouwenhoven, Science {\bf 281},
  540  (1998).

\bibitem{Park2002}
J. Park {\it et~al.}, Nature {\bf 417},  722  (2002).

\bibitem{Liang02}
W.~J. Liang {\it et~al.}, Nature {\bf 417},  725  (2002).

\bibitem{Elhassid00}
Y. Elhassid, Rev. Mod. Phys. {\bf 72},  895  (2000).

\bibitem{McEuen99}
M. Bockrath {\it et~al.}, Nature {\bf 397},  598  (1999).

\bibitem{Yacoby02}
O.~M. Auslaender {\it et~al.}, Sceince {\bf 295},  825  (2002).

\bibitem{Nitzan07}
M. Galperin, M.~A. Ratner, and A. Nitzan, J. Phys.: Condens. Matter {\bf 19},
  103201  (2007).

\bibitem{Ho98}
B.~C. Stipe, M.~A. Rezaei, and W. Ho, Science {\bf 280},  1732  (1998).

\bibitem{Jauho1994}
A.~P. Jauho, N.~S. Wingreen, and Y. Meir, Phys. Rev. B {\bf 50},  5528  (1994).

\bibitem{Schiller95}
A. Schiller and S. Hershfield, Phys. Rev. B {\bf 51},  12896  (1995).

\bibitem{Voit96}
Y.~P. Wang and J. Voit, Phys. Rev. Lett. {\bf 77},  4934  (1996).

\bibitem{Jauho_book}
H. Haug and A.~P. Jauho, {\em Quantum Kinetics in Transport and Optics of
  Semiconductors} (Springer, Germany, 1996).

\bibitem{Sugar81}
R. Blankenbecler, D.~J. Scalapino, and R.~L. Sugar, Phys. Rev. D {\bf 24},
  2278  (1981).

\bibitem{Hirsch88}
R.~M. Fye and J.~E. Hirsch, Phys. Rev. B {\bf 38},  433  (1988).

\bibitem{Flensberg03}
K. Flensberg, Phys. Rev. B {\bf 68},  205323  (2003).

\bibitem{Galperin2004}
M. Galperin, M.~A. Ratner, and A. Nitzan, J. Chem. Phys. {\bf 121},  11965
  (2004).

\bibitem{Mitra2004}
A. Mitra, I. Aleiner, and A.~J. Millis, Phys. Rev. B {\bf 69},  245302  (2004).

\bibitem{Hod06}
O. Hod, R. Baer, and E. Rabani, Phys. Rev. Lett. {\bf 97},  266803  (2006).

\bibitem{Muehlbacher01}
L. M{\"u}hlbacher, J. Ankerhold, and C. Escher, J. Chem. Phys. {\bf 121},
  12696  (2004).

\bibitem{Feynman63}
R.~P. Feynman and F.~L. {Vernon Jr.}, Ann. Phys. {\bf 24},  118  (1963).

\bibitem{Chen87}
Y.~C. Chen, J. Stat. Phys. {\bf 47},  17  (1987).

\bibitem{Orland_Negele}
J.~W. Negele and H. Orland, {\em Quantum Many-Particle Systems} (Perseus Books,
  Reading, MA, 1988).

\bibitem{Prokofev98}
N.~V. {Prokofe'ev}, B.~V. Svistunov, and I.~S. Tupitsyn, Phys. Lett. A {\bf
  238},  253  (1998).

\bibitem{Weissbook}
U. Weiss, {\em Quantum Dissipative Systems} (World Scientific, Singapore,
  1999).

\bibitem{Leggett87}
A.~J. Leggett {\it et~al.}, Rev. Mod. Phys. {\bf 59},  1  (1987).

\bibitem{Peskin03}
M. Caspary, L. Berman, and U. Peskin, Chem. Phys. Lett. {\bf 369},  232
  (2003).

\bibitem{Wingreen89}
N.~S. Wingreen, K.~W. Jacobsen, and J.~W. Wilkins, Phys. Rev. B {\bf 40},
  11834  (1989).

\end{thebibliography}

\end{document}